\documentclass{article}

\begin{document}
\title{On the fundamental structure of nature and the unification
of forces}
\author{J.Witters\\University of
Leuven,Department of Physics\\Celestijnenlaan 200D, 3001 Leuven,
Belgium} \maketitle
\begin{abstract}
A model for the fundamental structure of nature is presented. It
is based on two fundamental fermions moving with the velocity of
light and differing from each other by the projection of the spin
on the momentum vector. The energy of both fermions is
proportional to the momentum, which is scaled inversely with the
size of a length quantum. All the known forces are a manifestation
of one elementary interaction, the spin exchange or spin flip-flop
which takes place when two different elementary fermions come
together in the same space cell. At this stage the model can
explain the properties of the photon as a two-fermion particle and
it can be shown that the Dirac theory for relativistic fermions
could be deduced from this model. The model predicts that
particles like the electron or the quark are stable combinations
of a large number of the fundamental fermions, but proof of that
prediction has not been given.
\end {abstract}
\section* {introduction}
Today's theories and models of elementary particles and
interactions are based in part on ad hoc assumptions and have
internal inconsistencies which are unaccounted for. One of the
main difficulties is based on the fact that particles like
electrons or quarks, which have a relatively low rest energy,
appear in all experiments to be pointlike. The combination of low
energy and confinement to a very small region in space should be
impossible because basic laws tell us that confinement goes
together with high momentum components and that high momenta give
a high contribution to the energy. This difficulty arises also for
the photon which has zero rest mass and is so small in the plane
perpendicular to its momentum vector that it has virtually no
cross section for collisions with other photons.\\ The
contradiction between confinement and low energy can not be
discarded by an ad hoc introduction of negative potential energy
as is done for systems of interacting particles. Negative
potential energy is presented as resulting from the presence of
force-carrying particles like gravitons or photons or gluons.
These carriers of the interactions, if confined, should give
themselves a high momentum contribution to the energy. In a
consistent model all particles should be treated equal with
respect to the general laws of physics.
\\A second point which has not been treated satisfactorily is the
symmetry of negative and positive eigenvalues of the Dirac
Hamiltonian for relativistic fermions. The positive values have
been attributed to real particles whereas the negative ones are
piled up in an infinite sea of ghost particles and only the
defects in that sea of particles are thought to be real.\\The
model of the material world which is presented here deals with
those difficulties by introducing two fundamental fermions as
universal building blocks for all the particles, including those
which carry the interactions. The combination of high momenta with
low energy is made possible by accepting negative energy states as
real. Essential features of this model are a simplification of the
Dirac Hamiltonian by elimination of the rest mass term and a
reinterpretation of the Hamiltonian, making the square of it
responsible for the evolution of a system of interacting
particles.\section {the elementary fermions} To preserve the
antisymmetry which is manifested in so many ways in natural
phenomena, the fundamental building blocks should be fermions. The
uncertainty relation
\begin {equation} \Delta E \Delta t \geq \hbar\end {equation}
couples $\Delta E$, the uncertainty in the energy of a particle,
to the time span $\Delta t$ during which the energy is measured. A
similar relation holds between the uncertainty of the momentum p
and the dimension of the space cell in which it is evaluated. It
is therefore meaningless to define the value for the momentum or
the energy of a fundamental building block at one instant or at
one point in space. The closer a building block of a particle with
spherical symmetry comes to the center, the higher the value of
the azimuthal momentum component. The maximum value depends on the
choice of a quantum of space, conveniently introduced to avoid
infinities which result from taking r, the distance from the
center, to zero. If we choose a quantum $\Delta r_{o}$  and a
corresponding $\Delta t_{o}$ , related to it through  $\Delta
r_{o} = c \Delta t_{o}$  where c is the velocity of light, we set
the maximum momentum and energy of a fundamental building block
equal to $ p_{o} = \hbar /\Delta r_{o}$  and $ E_{o} = \hbar /
\Delta t_{o}.$\\ A spin $ 1/2$ fermion has two orthogonal spin
states. The spin up and the spin down variant can therefore be
different particles. They can be transformed into one another by a
spin flip-flop. If the transformation occurs every half cycle of a
cyclic motion, an averaged value of the momentum and the energy
can be calculated  by integration over the period of the cycle. It
is a well known phenomenon in atomic physics that in the presence
of strong coupling a new energy level halfway between two levels
is formed as a result of fast switching between these levels. A
quantity measured by integration over a period which is much
longer than the switching time is determined by the properties of
the new intermediate level. If this fast switching occurs between
a positive and a negative energy state, the intermediate level can
have an energy close to zero, even if the unperturbed energy
levels are far from zero.
\\The difference between this model and Dirac's theory for
elementary fermions, which also uses the option of positive and
negative energies, is that we start here with elementary fermions
moving at the velocity of light and therefore without rest mass,
and construct localised particles with them. In our model the rest
energy is a consequence of internal motion and it is very small
compared to the relativistic energies of the constituents. In
Dirac's theory the rest energy is introduced ad hoc. \\  We
introduce two elementary fermions called "ef" and use the name
particles for stable or quasi-stable combinations of efs. One ef
is righthanded (ref), the other one lefthanded (lef). They move
with the velocity of light and both have spin 1/2. For the ref the
momentum and spin vector are parallel and for the lef they are
antiparallel:\\$ref:momentum \rightarrow , spin
\rightarrow\\lef:momentum \rightarrow , spin \leftarrow$ \\

As will become clear by analysis of the velocity $d\vec{r}/dt$ ,
the direction of motion of each ef is along the spin vector rather
than the momentum vector. A ref and a lef "meet" each other by
entering the same space cell. In that cell the spin exchange leads
to a scattering of the  motion and this exchange is at the origin
of the fast ref-lef transitions which lower the energy of every
fundamental fermion bound in a particle. The fundamental
interaction between efs can be defined as a spin flip-flop. The
levels between which the fast switching takes place are $\pm
E_{o}$  for a particle with the size of an elementary space cell.
Combinations of equal numbers of  refs and lefs form bosons.
Fermionic particles are formed by combining n refs with m lefs
where m and n differ by an uneven number. This number difference
is probably only 1 for all the known fermions. Fermions in stable
form are therefore bosons combined with one extra ref or lef.
\section{short recapitulation of the Dirac theory for fermions}
 The Dirac Hamiltonian $H_{D}$  is a
four by four matrix of the form:\begin
{equation}H_{D}=c(\hat{\alpha}.\vec{p})+mc^{2}\hat{\beta}\end
{equation} where $\hat{\alpha}$ is a matrix composed of two Pauli
spin matrices $\hat{\sigma}$ in off-diagonal positions and $
\hat{\beta}$ has two unit matrices with different sign on the
diagonal:
\begin {equation} \hat{\alpha}=\left(\begin{array}{cc}
  0 &  \hat{\sigma}  \\
   \hat{\sigma} & 0
\end{array}\right) \end {equation}  and \begin
{equation} \hat{\beta}= \left(\begin{array}{cc}
  \hat{1} & 0 \\
  0 & -\hat{1}
\end{array}\right) \end {equation}

$ \vec{p} $  is the momentum operator $
-i\hbar\vec{grad}=-i\hbar\bar{\nabla} $ . The Pauli spin matrices
$ \sigma_{x} ,  \sigma_{y}, \sigma_{z} $ for spin quantisation in
the z-direction are given by:\begin {equation}
\sigma_{x}=\left(\begin{array}{cc}
  0 & 1  \\
  1 & 0
\end{array}\right) \end {equation}\begin {equation} \sigma_{y}=\left(\begin{array}{cc}
  0 &  -i  \\
  i & 0
\end{array}\right) \end {equation}\begin {equation} \sigma_{z}=\left(\begin{array}{cc}
  1 & 0  \\
  0 & -1
\end{array} \right)\end {equation}
$H_{D} $ operates on a four-vector as shown in the following
equation: \begin{equation} \small
\mathbf{H_{D}}=\left(\begin{array}{cccc}
  mc^{2} & 0 &-ic\hbar \frac{\partial}{\partial z} & -ic\hbar(\frac{\partial}{\partial x}-i\frac{\partial}{\partial y}) \\
  0 & mc^{2} &-ic\hbar ( \frac{\partial}{\partial x}+i\frac{\partial}{\partial y}) & ic\hbar\frac{\partial}{\partial z} \\
-ic\hbar   \frac{\partial}{\partial z}&-ic\hbar(
\frac{\partial}{\partial x}-i\frac{\partial}{\partial y}) &-mc^{2}
& 0 \\
  -ic\hbar(\frac{\partial}{\partial x}+i\frac{\partial}{\partial y}) & ic\hbar\frac{\partial}{\partial z} & 0 &
  -mc^{2}\end{array}\right)\\ \begin{array}{c}
    operating\\
    on \
  \end{array} \left(\begin{array}{c}
    f_{1}\mid+>\\
    g_{1}\mid-> \\
    f_{2}\mid+> \\
    g_{2}\mid-> \
  \end{array}
\right)
\end{equation} \normalsize
 In the four-vector on which the Hamiltonian
$ H_{D} $  operates $ f_{1}, g_{1}, f_{2}, g_{2} $  are functions
of space and time coordinates. This four-vector  can be
interpreted as two two-vectors $\phi $ and $\chi $ which are
transformed into each other and the eigenvalue $ \epsilon $ of
$H_{D} $ for an isolated particle with rest mass m can be obtained
from two matrix equations: \begin {equation} c
(\hat{\sigma}.\vec{p})\chi = (\epsilon - mc^{2})\phi
\end {equation}  \begin {equation} c (\hat{\sigma}.\vec{p})\phi=
(\epsilon + mc^{2})\chi
\end {equation}

The operator $ c (\hat{\sigma}.\vec{p}) $ is nondiagonal but the
square of this operator is diagonal with an eigenvalue  $
(\epsilon^{2} - (mc^{2})^{2}) $ which is equal to $ c^{2}p^{2} $ .
The eigenvalue $\epsilon$ is interpreted as the energy of the
particle. It can have a negative as well as a positive value, and
there is no a priori reason to prefer one over the other.
\section {the evolution operator in this model}
An operator acting exclusively on efs, which have no rest mass,
can have no $mc^{2}$ term as in Dirac's Hamiltonian.\\

The operator $ c (\hat{\sigma}.\vec{p}) $ ,based on Pauli spin
matrices  $ \hat{\sigma} $ is a Hamilton operator when it is
applied to isolated efs. Its plane wave eigenfunctions are:\\ $
Ne^{i\vec{k}.\vec{r}}\mid+> $ with eigenvalue $ c\hbar k $ \\ $
Ne^{i\vec{k}.\vec{r}}\mid-> $ with eigenvalue $ -c\hbar k $

where N stands for a normalisation factor and the spin eigenstates
$ \mid+> $ and $ \mid-> $   are for quantisation in the direction
of $ \vec{k} $.

All functions of space coordinates and the spin states for
interacting efs can be expanded in plane wave spinors. However,
for localised particles which are centered around one point or
around an axis, spherical or cylindrical spinors form a more
natural basis.\\The only type of spinor which is completely worked
out in this paper is that of the photon, with cylindrical
symmetry. Therefore the examples given below are for cylindrical
symmetry but the changes which have to be made for spherical
symmetry are straightforward.\\

In cylindrical coordinates the radius r is defined in the xy plane
\\\\
\setlength{\unitlength}{1mm}
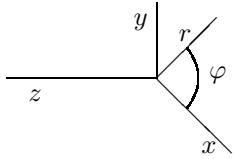
\begin{picture}(40,20)
\put(0,10){\line(1,0){20}} \put(20,10){\line(0,1){10}}
\put(20,10){\line(1,-1){10}} \put(20,10){\line(1,1){8}}
\qbezier(24,14)(27,10)(24,6) \put(3,7){$z$}\put(26,0){$x$}
\put(17,17){$y$}\put(23,15){$r$}\put(27,10){$\varphi$}
\end{picture} \\\\

 $ c (\hat{\sigma}.\vec{p}) $ can be conveniently written in terms of raising and lowering operators:
\begin{equation}
c\hat{\sigma}.\vec{p}=-i\hbar
c(\sigma_{z}\partial_{z}+\sigma_{+}\partial_{-}+\sigma_{-}\partial_{+})
\end{equation}
with
\begin{equation}
\sigma_{\pm}=\frac{\sigma_{x}\pm\sigma_{y}}{\sqrt{2}}
\end{equation}
and
\begin{equation}
\partial_{\pm}=e^{\pm
i\varphi}(\frac{\partial}{\partial_{r}}\pm\frac{i}{r}\frac{\partial}{\partial\varphi})
\end{equation}
 The spinor
wavefunction $\psi$ of a particle with two or more efs can be
written as a symmetrized combination of products of spinors, each
spinor referring to the coordinates of one ef:
\begin{equation}
\mathbf{\psi}=\left(\begin{array}{c}
  u_{1} \\
  v_{1}
\end{array}\right)\left( \begin{array}{c}
  u_{2} \\
  v_{2}
\end{array}\right)\ldots \left(\begin{array}{c}
  u_{i} \\
  v_{i}
\end{array}\right)+symmetric\: permutations\end{equation}
where the $u_{i}$  and the $v_{i}$ are fuctions of space and time
coordinates. \\The operator $ c (\hat{\sigma}.\vec{p}) $   , which
we will denote by $H$, although it is in general not diagonal and
therefore not an Hamiltonian, must act consecutively on each of
the spinors and could therefore be represented as a sum:
\begin{equation}
H=\sum   c (\hat{\sigma}_{i}.\vec{p}_{i})
\end{equation}
The sum should run over all efs which are not excluded by
antisymmetry from the space occupied by the particle. It should
therefore include efs in vacuum states when they interact with the
efs in a particle. $H$ is a local operator and so is also
$H^{*}H=-H^{2}$ . The ij product terms in $ H^{2}$ refer to
different efs when $ i \neq j $ but the interaction must be local,
referring to efs in the same space-time cell.\\ \large $H^{2}$ is
the interaction operator in our model. \\ The combination of
raising and lowering which transforms a lef in a ref and vice
versa, is induced by $ H^{2}$ , more specifically in  the terms
producing a spin exchange in $ \mid +> \mid ->  or \mid->\mid +>$
combinations.\normalsize \\ The spinor function of a stable
composite particle for which the lef-ref transformations form
closed vertices must therefore be an eigenfunction of $ H^{2}$.
The "Zitterbewegung" which was recognised shortly after the
introduction of the Dirac Hamiltonian is a consequence of this
cyclic raising and lowering. It is easily interpreted in our model
because $d\vec{r}/dt$ can be evaluated by commuting $\vec{r}$ with
$ c (\hat{\sigma}.\vec{p}) $ , the result of this commutator being
$c\hat{\sigma}$ . A bound ef executes a "Zitterbewegung", a sort
of spiraling motion which is guided by the spin precession.

For non-interacting efs the cross terms in $H^{2}$ are absent and
then we have, by virtue of
\begin{equation}
(\bar{\sigma}\cdot\bar{A})(\bar{\sigma}\cdot\bar{B})=\bar{A}\cdot\bar{B}+i\bar{\sigma}\cdot(\bar{A}\times\bar{B})
\end{equation}:
\begin{equation}< HH > = < \sum (c \hat{\sigma}_{i}.\vec{p}_{i})( c
\hat{\sigma}_{i}.\vec{p}_{i})>= -\sum c^{2}p_{i}^{2}
\end{equation}

where the $ p_{i} $ are numbers and $<...>$ stands for integration
in order to calculate the expectation value of the operator.
\\ Then the operator $ H^{2} $ measures minus the sum of the
squares of the energies. The interaction terms give contributions
which reduce the norm of $<  H^{2}> $  , the expectation value of
$ H^{2} $.\\ Efs are never really isolated. Even if they are
subjected to strong localisation in particles the tails of the
spinors will extend in space and overlap with other particles. The
exchange of photons, which are described further on, is a form of
interaction through vacuum polarisation and is much more efficient
than an overlap by exponentially decaying tails. The products of
spinors in the expression for the total wavefunction which
describes several interacting particles should always be completed
by including all the photons  allowed by the available degrees of
freedom. The interaction may reduce $< -H^{2}> $  , by increasing
the overlap for a ref-lef combination or decreasing it for ref-ref
or lef-lef combinations. It would leave $< H^{2}> $ unchanged if
the particle would be isolated and in an eigenstate of $ H^{2}$.
But this is never realised.  Any change in $< H^{2}> $ for the
initial system of efs will give rise to the emission of new
particles and thereby increase the entropy of the system, and this
process goes on continuously.
\\ The evolution of a system of particles is a consequence of a
tendency towards maximum entropy.
\\ The physical interpretation of the $  H^{2} $ operator which
emerges from these considerations is that it represents the
negative square of the energy if applied to one particle and minus
the sum of the squares of energies for several non-interacting
particles. Creation of new particles, a process which goes on
continuously, reduces $< - H^{2}> $ because the new particles are
formed under conservation of total momentum and the sum of the
squares of fractions is smaller than the square of the whole. In
this process the entropy of the system is increased. $< -H^{2}> $
is therefore never really constant. It may be approximately
constant for a given restricted time scale and for corresponding
restricted regions of space, in which all the efs execute cyclic
motions, being bound in well-defined particles. This corresponds
to having a set of closed vertices in Feynman diagrams.  $< -
H^{2}> $   changes when the vertices are opened up and new
particles are formed.\\ A proposition which is essential to our
model is that a fermionic particle, consisting of an ef with a
cloud of refs and lefs in equal numbers bound to it, can interact
with a similar particle and form a new quasi-stable eigenstate of
$ H^{2} $ . In that case the particles formed in the interaction
are virtual particles, emitted but also reabsorbed in a closed
interaction vertex. This is at the origin of the formation of
structures like the proton with three fermionic particles or the
hydrogen atom with a proton interacting with an electron and so
on.
\section{the link with the Dirac Hamiltonian}
For a particle with the center of mass c.m. at a position $
\vec{r}_{m} $ and momentum  $ \vec{p}_{m} $ corresponding to the
motion of the c.m., and with internal position vector  $
\vec{r}_{i} $  and corresponding momentum  $ \vec{p}_{i} $ for the
efs in the particle, the total momentum operator can be expressed
as a sum $ \vec{p}_{m} +   \vec{p}_{i} . $  The operator $ H $ can
consequently be split in two parts:
\begin{equation}
H= \sum c \hat{\sigma}_{i}.\vec{p}_{i} +  \sum c
\hat{\sigma}_{i}.\vec{p}_{m}
\end{equation}

The first part refers to an internal contribution  to the energy
of the particle and can therefore be identified with the rest
energy $ mc^{2}. $   The second part refers to the motion of the
c.m. and is the kinetic term. The rest energy term is zero if the
contributions of refs and lefs cancel out. It will be shown that
this is the case for the photon in our model, and it can be
expected from symmetry arguments that the same conclusion will
hold for all single- particle bosons in the ground state. \\ For
fermionic particles with only a few refs and lefs the lowest
possible value for the rest mass term will be huge on account of
the huge value of each $ p_{i}.$ Only when large numbers of efs
combine together and lead to almost complete cancellation will the
rest mass term be relatively small with respect to the energies of
the efs. With respect to the energy in the c.m. motion however,
the rest mass term of the particles studied in the laboratory is
very big in most experiments. That is the typical situation to
which the Dirac theory is applied. \\ The Dirac Hamiltonian shown
in equation (8) hides the internal coordinates of a fermionic
particle. The transformation from positive to negative energy
states refers to the elementary interaction with the raising and
lowering operators which are combined with spin flips of the efs.
Each elementary interaction causes a switching between these
states. The total spin of the particle however may or may not be
changed by the interaction.

 $ \partial /\partial z $ transforms $ f_{1} $ in (8) into $
f_{2} $ and leaves the spin unchanged, whereas $
\partial /\partial x + i\partial / \partial y $
transforms $ f_{1} $ into $ g_{2} $ with a spin-flip and $
\partial / \partial x - i\partial / \partial y $ transforms
$ g_{1} $ into $ f_{2} $ with a reverse spin-flip. The spins refer
here to the total fermion, not to the individual efs in it. Some
non-diagonal terms in the Dirac Hamiltonian, non-diagonal terms
which are responsible for the coupling between positive and
negative rest energies, cause spin flips whereas other terms leave
the spin unchanged. Therefore the sign of the scalar spin-momentum
product and the sign of the energy may be different for the states
described by a four-vector, in contrast with efs for whitch these
signs are identical. \\A fermionic particle therefore has
different quantum numbers for the spin and the charge, the latter
 discriminating between particles and antiparticles. \\An ef on the contrary can be
caracterised by one quantum number, the sign of the  projection of
the spin on the momentum vector.\\ A particle has the same charge
as the extra ef in it and differs from the corresponding
antiparticle by the fact that one has a ref and the other a lef
combined with an uncharged cloud.
\section{the photon}
In its ground state the photon has zero energy and it must consist
of an equal number of refs and lefs in very strong interaction,
giving a complete cancellation of + and - terms in the energy.
Excitation gives the photon an energy equal to c times its
momentum in the direction of propagation. When a photon is
exchanged between particles, the total angular momentum, orbital
plus spin, which is transmitted is equal to $ \hbar. $   The
photon exhibits cylindrical symmetry, being delocalised in the
direction of its angular momentum. It is therefore practical to
use cylindrical spinor functions to describe the wave function of
the photon. These spinors must be eigenfunctions of $ H^{2} $ ,
and therefore also of the $ \bar{\nabla}^{2} $ operator in
cylindrical coordinates.This operator is:
\begin{equation}
\left(\frac{\partial}{\partial z}\right)^{2}+
\partial_{+}\partial_{-}=\left(\frac{\partial}{\partial
z}\right)^{2}+\left(\frac{\partial}{\partial
r}\right)^{2}+\frac{1}{r}\frac{\partial}{\partial r}+
(\frac{1}{r})^{2} \left(\frac{\partial}{\partial
\varphi}\right)^{2}
\end{equation}

The following cylindrical spinor function $\zeta$   is an
eigenfuction with  angular momentum $ \hbar $ , non-singular for
r=0, and zero momentum in the z-direction:\\
    Spinor $\zeta$  :   $  u(r,\varphi )\mid + > +  v(r,\varphi )\mid -
    > $\\
where r and $ \varphi $ are cylindrical coordinates, the spin
states are quantised in the z-direction and u and v are given by:

\begin{equation}
u(r,\varphi)=N\frac{sin(k_{o}r)}{\sqrt{k_{o}r}}e^{i\frac{\varphi}{2}}
\end{equation}
\begin{equation}
v(r,\varphi)=N\left(\frac{cos(k_{o}r)}{\sqrt{k_{o}r}}-\frac{sin(k_{o}r)}{\sqrt{(k_{o}r)^{3}}}\right)e^{3i\frac{\varphi}{2}}
\end{equation}
It is interesting to note that in cylindrical symmetry the orbital
angular momenta are formed by half odd integer quantum numbers, as
opposed to integers for spherical symmetry.\\ The separate
transformations of the raising and lowering operators are:
\begin{equation}
\sigma_{+}\partial_{-} v(r,\varphi) \mid-> = -k_{o}u
(r,\varphi)\mid+> \end{equation}
\begin{equation}
\sigma_{-}\partial_{+} u(r,\varphi) \mid+> = k_{o}v
(r,\varphi)\mid-> \end{equation} The spinor function $\zeta$  can
be interpreted either as a stable bosonic state with total angular
momentum composed of a mixture of orbital- and spin contributions
or as a function describing one fermion which alternates between
ref and lef so quickly that it loses identity. The alternation is
caused by spin tumbling through interaction with another ef with
opposite spin and momentum, whose wave function  $\xi$   is:

     Spinor $\xi$  :   $  \acute{u}(r,\varphi )\mid - > +  \acute{v}(r,\varphi )\mid
     +
    > $
\begin{equation}
\acute{u}(r,\varphi)=N\frac{sin(k_{o}r)}{\sqrt{k_{o}r}}e^{-i\frac{\varphi}{2}}
\end{equation}
\begin{equation}
\acute{v}(r,\varphi)=N\left(\frac{cos(k_{o}r)}{\sqrt{k_{o}r}}-\frac{sin(k_{o}r)}{\sqrt{(k_{o}r)^{3}}}\right)e^{-3i\frac{\varphi}{2}}
\end{equation}
In  diagrammatic form the photon could be described as a string of
closed vertices, the interaction points referring to the spin
flip-flops, as shown in the next picture:
\\\\
\setlength{\unitlength}{1mm}
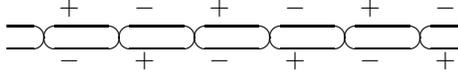
\begin{picture}(60,20)
\put(-5,10){\oval(10,3)[r]} \put(5,10){\oval(10,3)}
\put(15,10){\oval(10,3)} \put(25,10){\oval(10,3)}
 \put(35,10){\oval(10,3)}\put(45,10){\oval(10,3)}
\put(55,10){\oval(10,3)[l]}\put(2,13){$+$}\put(2,6){$-$}\put(12,13){$-$}\put(12,6){$+$}\put(22,13){$+$}\put(22,6){$-$}\put(32,13){$-$}\put(32,6){$+$}\put(42,13){$+$}\put(42,6){$-$}\put(52,13){$-$}\put(52,6){$+$}
\end{picture} \\\\
Each ef changes sign regularly and the energy is zero when
integrated over a cycle. One single photon in a number state has a
ref and a lef, with opposite spin but in an undefined phase of the
cycle. Its spinor function is a combination of the up part of $
\zeta$  with the down part of $\xi$   and vice versa:
\begin{equation}
\Psi_{zero\: energy\:
photon}=u(1)\mid+>\acute{u}(2)\mid->+v(1)\mid->\acute{v}(2)\mid+>
\end{equation}

where (1) and (2) refer to the spatial coordinates of the first
respectively the second ef and the spin states are coupled to the
preceding wavefunction. For this spinor combination the energy,
the total momentum and the total angular momentum are zero. It is
a spinor in the "vacuum". To transform it into a real photon which
transmits energy the $\zeta$   and the $\xi$  part in it must have
a momentum component in the z-direction:
\begin{equation}
\Psi_{photon}=e^{ikz_{1}}u(1)\mid+>e^{-ikz_{2}}\acute{u}(2)\mid->+e^{ikz_{1}}v(1)\mid->e^{-ikz_{2}}\acute{v}(2)\mid+>
\end{equation}

transmitting an energy  $\hbar k c$ and an angular momentum
$\hbar$ . Source-detector exchange of a photon is a coupling of
the source to the detector by a diagram as shown above. The source
and the detector interact with the same photon, each on one end.
When the source interacts with the $\zeta$ part, the detector
takes the $\xi$  part and vice versa. This is shown symbolically
in the next figure:
\\\\
\setlength{\unitlength}{1mm}
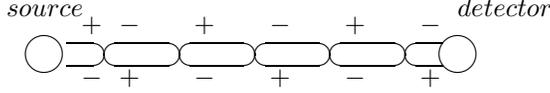
\begin{picture}(60,20)
\put(2,10){\circle{5}} \put(5,10){\oval(10,3)[r]}
\put(15,10){\oval(10,3)} \put(25,10){\oval(10,3)}
 \put(35,10){\oval(10,3)}\put(45,10){\oval(10,3)}
\put(55,10){\oval(10,3)[l]}\put(7,13){$+$}\put(7,6){$-$}\put(12,13){$-$}\put(12,6){$+$}\put(22,13){$+$}\put(22,6){$-$}\put(32,13){$-$}\put(32,6){$+$}\put(42,13){$+$}\put(42,6){$-$}\put(52,13){$-$}\put(52,6){$+$}
\put(57,10){\circle{5}}\put(-3,15){$source$}\put(57,15){$detector$}\end{picture}
\\\\

For the photon the transmission event is simultaneous on both
sides, the time being relativistically contracted. In this picture
there is no "strange correlation at a distance", an expression
which is frequently used in the literature when quantum
correlations are examined. The detector and the source are on
equal terms here and there is nothing strange about the fact that
the source "knows" which detector, eventually far away from it, is
going to be selected.\\

The appearance of half-integer quantum numbers for the orbital
angular momenta of the spinors is not in contradiction with the
bosonic symmetry of a photon. The spin is rotating synchronously
with the orbiting ef and one turn over the orbit must therefore
change the sign of the wavefunction, this change in sign being
compensated by the spin rotation. The transformation formulae for
change of spin-quantisation axis, from parallel to the momentum ($
\mid >_{p}$ ) to parallel to z ($\mid > $ ) are given by:
\begin{eqnarray}
 \mid + >_{p} = e^{-i\frac{\varphi}{2}} cos
(\frac{\theta}{2})\mid +> + e^{i\frac{\varphi}{2}} sin
(\frac{\theta}{2})\mid ->\\ \mid->_{p} =- e^{-i\frac{\varphi}{2}}
sin (\frac{\theta}{2})\mid +> + e^{i\frac{\varphi}{2}} cos
(\frac{\theta}{2})\mid ->\end{eqnarray}

where $\theta$  and $\varphi$  are the polar, resp. azimuthal
angles of the spin vector with respect to the z-axis. For $\theta
= \frac{\pi}{2}$ , as in the cylindrical frame, and for the spin
of an ef rotating in the xy plane with an azimuthal part of the
wavefunction $ exp(i\varphi )$ , we obtain a combination of
azimuthal functions as in the spinor $\zeta$ . Rotation described
by $ exp(-i\varphi )$ leads to spinor $\xi$  .\\ The zero energy
photonic spinors introduced above can fill up all the available
space in the vacuum. This representation of the vacuum is an
alternative for a representation with plane wave spinors, but is
better suited to treat the interaction between two particles. A
system of particles can only be in a stable configuration if all
interactions with the photons which are allowed in the region of
space occupied by the system lead to closed vertices, i.e. to
cyclic motion of all the efs in the system.

Eigenstates of the $ \bar{\nabla^{2}}$  operator in spherical
symmetry are the spherical Bessel functions, quite similar to the
functions in the cylindrical spinors but with integer rather than
half integer powers of $k_{o}r$  in the denominators, the lowest
order function being $sin(k_{o}r )/k_{o}r$ .
\section{fermions}
Combinations of ref-lef pairs and one extra ref or one extra lef
which form stable localised particles are fermions. This follows
from the rule for summing up an odd number of half integer spins
and the fact that localised particles with spherical symmetry have
an integer quantum number for the orbital angular momentum. \\For
a ref-lef-ref trio in a c.m. frame the three momentum vectors add
up to zero, as shown in the next figure:
\\\\
\setlength{\unitlength}{1mm}
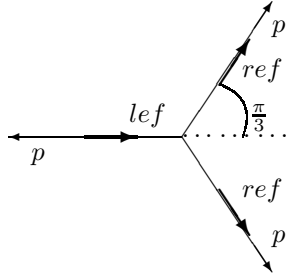
\begin{picture}(80,40)
\put(40,20){\ldots}\put(45,20){\ldots}\put(50,20){\ldots}\qbezier(45,27)(50,25)(48,20)
\put(40,20){\vector(-1,0){23}}\put(40,20){\vector(2,3){12}}
\put(40,20){\vector(2,-3){12}}\put(20,17){$p$}\put(52,34){$p$}\put(49,22){$\frac{\pi}{3}$}
\put(52,6){$p$}\thicklines\put(27,20){\vector(1,0){7}}\put(45,27){\vector(2,3){4}}
\put(45,13){\vector(2,-3){4}} \put(33,22){$lef$}\put(48,28){$ref$}
\put(48,12){$ref$}
\end{picture} \\\\

Projection of the momenta leads to: $- p cos(0) + 2 p
cos(\frac{\pi}{3}) = 0 $ . Projection of the spin vectors give
components with cosinus and sinus functions of half angles. The
combined spin of a ref-lef duo can not sum up with the spin of the
remaining ref to a  zero spin state. \\A simple argument, based on
the idea that the extra ref can interact with the photon-like
ref-lef combination which has in itself acquired an energy of half
the energy of an ef indicates that the resulting $<
\hat{\sigma}\cdot \bar{\nabla}> $  might be reduced by a factor
three.This would still mean a huge rest energy for the ref-lef-ref
trio.\\ As more efs are combined in one particle, the space needed
to preserve antisymmetry, that is for the efs of the same type to
avoid one another, must be larger. By analogy with examples in
atomic and nuclear physics a "closed shell effect" can be
expected, giving the largest energy reduction for efs in  a closed
packing within a certain radius. Starting with a given number of
efs in interaction but not in the lowest energy state, the
dynamics leading to a stable particle will involve the evaporation
of photons or other particles. In this process the total energy of
the particle plus the photons is constant and the dynamics are
therefore not dictated by a tendency to a minimum energy.\\ \large
The number of entities contributing to the energy, and through
that number to the entropy of the system, is the driving factor
for the evolution towards stable localised structures.\\
\normalsize Particles like the electron, with a very low rest
energy, must consist of a very large number of efs. A rough
estimate of the number n could be based on the consideration that
the ref-lef-ref trio of the example given above can in turn be
treated as a particle with momentum smaller than p, and coupled to
new efs, and so on. The trend of the energy as a function of n
should then follow some power law like $\kappa^{n} c p_{o}$ ,
where $\kappa$ is smaller than 1. For  $\kappa =\frac{1}{2}$ and
n=100 the reduction would be of order $10^{-30}$ with respect to
$cp_{o}$. \\ In the framework of this model a proton should
consist of three particles of which one has a negative charge like
the electron and two are positive like the positron and all have
spin $\frac{1}{2}$. The Pauli exclusion  principle is not violated
by considering two fermions with the same charge and spin in close
interaction because we are dealing with composite particles which
need not be identical. They might differ by even numbers of efs
which are exchanged between them. The forces binding the particles
in a nucleus may be strong enough to disrupt the internal
structure of the particles and cause massive inter-particle
transfers of efs, giving rise to the weak and the strong nuclear
interactions.
\\To include also neutrons and neutrinos in the model we must
explain a combination of spin $\frac{1}{2}$ with zero charge. We
identify the charge sign with the spin of the extra ef, taking one
sign for the ref and the opposite sign for the lef. A fermion
should therefore have a fixed charge on a very short time scale.
However, although the sign of the charge is preserved in a spin
flip-flop between two efs , a collective interaction involving all
the efs in a particle and inducing a turnover of all the spins at
once will change that sign. If this is repeated periodically and
with a very short period, all measurements of electric interaction
by integration over a longer period will show zero charge.
Neutrinos could therefore be fermions in a state of rapid
collective turnover from ref to lef character. Neutrons could
differ from protons by having one fermion with the same rapid
turnover.\\ Charge conjugation which transforms matter into
antimatter is identified here with ref-lef transition. Given that
particles are composed of ref-lef combinations, matter and
antimatter are intimately mixed in all structures. The asymmetry
in atoms here in our solar system, with all nuclei positive and
all electron clouds negative, must follow from an initial
fluctuation at the moment of formation of these
structures.\section{electromagnetism and gravitation} Interaction
between two particles, in the limit of large distance between the
centers, can be treated by an approximation in which the energy of
the separate particles is determined independently, and
subsequently the overlap of the spinors from the two centers and
the connection through photons from the vacuum is considered. We
can try to identify the two well-known long distance interactions:
electromagnetism and gravitation. The first is caused by
interparticle correlation of the extra efs in each particle, the
extra efs carrying the charge of the particle. The interaction
between these extra efs is mediated through flip-flop interactions
with the photon spinors in the vacuum and the interaction strength
is proportional to the momentum of the extra efs in the
interacting particles. The second one is due to the correlation of
the total ref-lef unbalance in each particle, unbalance which is
expressed in the rest energy plus eventually the kinetic energy of
the c.m. of each particle. This is independent of the charge sign
and can be identified with gravitation. In the argument which was
used to explain the reduction in energy in the ref-lef-ref trio a
gravitational type of interaction was invoked. Calculation of the
interaction strength between quasi-stable particles will be a
formidable task, given that a very large number of efs moving
under conditions of strong correlation are involved. In a crude
model as indicated in the next figure the order of magnitude of
the electromagnetic correction can be estimated:

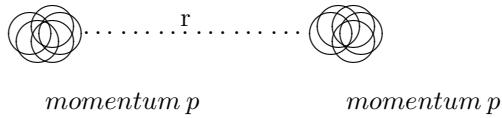
\setlength{\unitlength}{1mm}
\begin{picture}(80,35)
\put(22,20){\circle{6}}\put(18,20){\circle{6}}\put(21,21){\circle{6}}
\put(19,19){\circle{6}}\put(21,19){\circle{6}}\put(62,20){\circle{6}}\put(58,20){\circle{6}}\put(61,21){\circle{6}}\put(59,21){\circle{6}}
\put(61,19){\circle{6}}\put(25,20){\ldots}\put(30,20){\ldots}\put(35,20){\ldots}\put(40,20){\ldots}\put(45,20){\ldots}\put(50,20){\ldots}\put(38,21){r}\put(20,10){$momentum\:
p$}\put(60,10){$momentum\: p$}
\end{picture}

The largest value of the momentum p of efs in the interacting
particles depends on the size of the particles. The sizes of both
particles  are taken to be equal. We identify $cpr_{o} / r$, the
energy correction at distance r obtained by switching the momentum
p of the extra ef, with the electric energy$ e^{2} / 4\pi \epsilon
_{o}r$  and obtain $ pr_{o} = \frac{\hbar}{137}$. The reduction by
137 of the momentum p with respect to the maximum $p_{o}= \hbar /
r_{o}$ might be an indication of the increased wavelength in a
larger particle. 137 could be related to the number of efs in a
typical particle like the electron. It might correspond to the
number of cells, densely packed up for instance to the third
nearest neighbour.\\ To derive an order of magnitude for $p_{o}$
we could take the gravitational interaction to the limit of
interaction between two neighbouring efs and identify $c\hbar /r$
with $G (p_{o} / c)^{2} / r$, where G is the universal constant
for gravitation and the masses have been replaced by the full
kinetic masses $ p_{o} / c$. The value for $cp_{o}$ comes out as
$2.10^{9}$ Joule or $\approx 10^{28}$ eV, a huge value
corresponding to $r_{o}$ of the order of $10^{-35}$ meters. This
makes the cross section of the photon so small that it escapes
experimental verification.
\\\\\\ \large Concluding remarks:\normalsize \\ The model proposed
here puts particles and force carrying particles in one category.
It gives a simple and universal principle for all interactions:
the spin flip-flop or exchange, which occurs when elementary
fermions enter the same space cell. Elementary fermions have no
rest mass and therefore the operator for the energy depends  only
on spin and momentum operators. Dirac's relativistic Hamiltonian
for fermions is a special case when composite one-center particles
are considered and only the coordinates of the center of mass
together with the total angular momentum components are taken as
variables. The model can be applied successfully to explain the
combination of angular momentum $\hbar$ and zero rest mass and
zero charge which we know as the photon. The photon is the only
particle which has been completely described in this model. An
estimate of its cross-section is so small that it escapes
experimental verification. All this gives strong arguments for
trying to transform this model into a theory by calculating the
ground state energy of a fermionic particle like the electron.
\\\\ Final note by the author: \\ This paper was presented in slightly different form for
publication to Annalen der Physik in july 1998 . The editors
replied after a few months that two reviewers had left it without
comment and that a third reviewer would be contacted. There was no
further message up to now.
\end {document}